\documentclass[conference]{IEEEtran}

\usepackage{color}
\usepackage{subfigure}
\usepackage{graphicx}
\usepackage{amsmath}
\usepackage{amssymb}
\usepackage{wrapfig}
\usepackage{algorithmic, algorithm}
\usepackage[T1]{fontenc}
\usepackage{multirow}
\usepackage{paralist}
\usepackage{ulem}
\usepackage{amsthm}
\usepackage{url}
\usepackage{array}

\begin{document}

\title{Joint On-the-Fly Network Coding/Video Quality Adaptation for Real-Time Delivery}

\author{
\IEEEauthorblockN{Tuan Tran Thai, J\'er\^ome Lacan, Emmanuel Lochin}
\IEEEauthorblockA{University of Toulouse; ISAE/DMIA; T\'eSA; Toulouse, France}
}

\maketitle

\begin{abstract}
This paper introduces a redundancy adaptation algorithm for an on-the-fly erasure network coding scheme called Tetrys in the context of real-time video transmission. The algorithm exploits the relationship between the redundancy ratio used by Tetrys and the gain or loss in encoding bit rate from changing a video quality parameter called the Quantization Parameter (QP). Our evaluations show that with equal or less bandwidth occupation, the video protected by Tetrys with redundancy adaptation algorithm obtains a PSNR gain up to or more 4$\,$dB compared to the video without Tetrys protection. We demonstrate that the Tetrys redundancy adaptation algorithm performs well with the variations of both loss pattern and delay induced by the networks. We also show that Tetrys with the redundancy adaptation algorithm outperforms FEC with and without redundancy adaptation.
\end{abstract}

\IEEEpeerreviewmaketitle

\section{Introduction}

Video traffic currently plays an important role on the Internet. The delivery of multimedia content has been intensively researched to provide better service and quality to end users. H.264/AVC (Advanced Video Coding), video coding standardized since 2003, has shown better compression performance than previous standard codecs such as MPEG-4 Part 2, H.263 \cite{WiegandSBL03}. Additionally, the newly standardized video codec, High Efficiency Video Coding (HEVC) \cite{hevc} provides up to 50\% bit rate savings for equivalent perceptual quality compared to H.264/AVC. However, the higher compression efficiency makes the encoded video more sensitive to errors and losses during transmission on networks. A small number of losses can significantly degrade the video quality perceived by end users. Thus, the challenge in real-time video transmission over error prone networks is twofold:

\begin{enumerate}
\item Video traffic must be protected from losses over the Internet. Indeed, Wenger \cite{Wenger03} showed that the Peak Signal to Noise Ratio (PSNR) decreases up to several dB when the loss rate is greater than 1\%. From the video perspective, error resilience tools \cite{Wenger03, KumarXMP06} (e.g., data partition, Flexible Macroblock Ordering) provided by the video codec standards are designed to mitigate the impact of packet loss. However, these tools usually use extra bit rate which leads to lower coding efficiency \cite{h264evaluation}. From the network perspective, the obvious way to provide reliability is retransmission. Nevertheless, the delay to recover the lost packets requires at least one additional Round Trip Time (RTT) which is not applicable for interactive applications. The traditional approach is to use Forward Error Correction (FEC) \cite{rfc5510} to protect the video from losses. The main problem of this block code scheme is that it requires dynamically adapting its initial parameters and as a result, complex probing and network feedback analysis. Recently, novel erasure network coding approaches that prevent such complex configuration have been proposed \cite{arq_nc,Lacan,Tetrys}. The main difference between these proposals is that the code in \cite{Lacan,Tetrys}, called Tetrys, is more suitable for real-time video applications as this code is systematic and the repair packets in \cite{Lacan,Tetrys} are equally distributed between data packets.

\item The network condition (e.g., delay, loss rate) varies over the time. Hence, it requires an adaptive mechanism for erasure codes to adapt to network dynamics. In \cite{RED-FEC}, the authors propose a Random Early Detection FEC mechanism in the context of video transmission over wireless networks. This mechanism adds more redundancy packets if the queue at the Access Point is less occupied and vice versa. However, this approach assumes that the wired segment of the network is loss free. In reality, the wired segment of the network might experience packet losses due to cross traffic or network congestion. The approach in \cite{Lamoriniere} switches between different FEC techniques to adapt to the state of the network in the context of multi-source streaming. 
\end{enumerate}

Sahai in \cite{Sahai} showed the more the redundancy introduced on the network the shorter the packet recovery delay. Tetrys exhibits the same behavior for the stationary channel \cite{Tetrys}. However, when the channel state varies over the time, it is more difficult to control the variations of the redundancy ratio. Thus, in this article, we propose a redundancy adaptation algorithm for Tetrys to cope with network dynamics (e.g., loss rate and delay variations) in the context of real-time video transmission. Our algorithm adapts the Tetrys redundancy ratio by increasing or decreasing the video quality in order to deliver video in which the residual packet loss rate is minimized as much as possible within the delay constraint required by the application. Indeed, the algorithm exploits the relationship between the gain or loss in the encoding video bit rate by changing the video quality parameter, the Quantization Parameter, and the redundancy ratio used by Tetrys. From experiments with both x264 \cite{x264} and JM \cite{jm} video encoders, we observe that each time the QP varies by one, the gain (or loss) in terms of encoding bit rate ranges from 10\% to 20\% while the video quality PSNR decreases (or increases) in range from 0.5 to $1\,$dB. The slightly degraded video up to $2\,$dB caused by the QP increase does not significantly interfere with the visual impact experienced by end users while the degraded video caused by unrecovered lost packets or late-arrival packets has a significant visual impact. Furthermore, we chose the Tetrys redundancy ratio list so that the video with slightly lower quality protected by Tetrys does not send more bit rate than the video with higher quality but without protection from the erasure codes. The results show that Tetrys with redundancy adaptation algorithm gains on average up to or more than $1\,$dB compared to Tetrys without redundancy adaptation algorithm and more than $4\,$dB compared to the video without protection. It is noted that the subjective evaluation from watching the resulting videos shows much better perceived quality obtained by Tetrys with redundancy adaptation algorithm compared to Tetrys without redundancy adaptation algorithm \cite{tetrys_website}. The simulation results show that the algorithm adapts well to both loss pattern and delay induced by networks. We also show that Tetrys with redundancy adaptation algorithm outperforms FEC with and without redundancy adaptation algorithm.

The rest of this article is organized as follows: Section \ref{sec:tetrys_overview} briefly introduces the principle of Tetrys and notes some important properties. The redundancy adaptation with Tetrys is described in detail in Section \ref{sec:adaptive_tetrys}. Section \ref{sec:correlation} presents the rationale behind a chosen redundancy list for H.264/AVC real-time transmission. Section \ref{sec:evaluation_cbr} studies the impacts of algorithm parameters using Constant Bit Rate traffic. The evaluation with video traffic is presented in Section \ref{sec:evaluation_video}. Tetrys compared with FEC is the topic of Section \ref{sec:comparison_fec}. Section \ref{sec:related_work} discusses the differences between our approach and existing work. Concluding remarks are given in \ref{sec:conclusion}.

\section{Tetrys overview} 
\label{sec:tetrys_overview}

Tetrys \cite{Tetrys} is an erasure network coding scheme that uses an elastic encoding window buffer $B_{EW}$. This buffer stores all source packets sent and not yet acknowledged. For every $k$ source packets, Tetrys sender sends a repair packet $R_{(i..j)}$ which is built as a linear combination of all packets currently in $B_{EW}$ from packets indexed \textit{i} to \textit{j} 

$$R_{(i..j)} = \sum_{l=i}^{j}  \alpha^{(i,j)}_l.P_l$$

\noindent where the coefficients $\alpha^{(i,j)}_l$ are randomly chosen in the finite field $\mathbb{F}_{q}$. Through this coding, the redundancy ratio is specified as $1/(k+1)$ or $1/n$ (where $n=k+1$) which is equivalent to the code rate $k/(k+1)$. Unlike TCP that acknowledges every received packet, Tetrys receiver is only expected to periodically acknowledge the received or source decoded packets. Upon reception of acknowledgment packet, Tetrys sender removes the acknowledged source packets out of its $B_{EW}$. Generally, Tetrys receiver can decode all lost packets as soon as the number of received repair packets is equal to the number of lost packets. By this principle, Tetrys is tolerant to burstiness losses in both source, repair and acknowledgment packets as long as the redundancy ratio exceeds the packet loss rate (PLR). Furthermore, the lost packets are recovered within a delay that does not depend on the Round Trip Time (RTT). This property is very important for real-time applications where the time constraint is stringent. 

\begin{figure}[!htb]
\begin{center}
\includegraphics[scale=0.40]{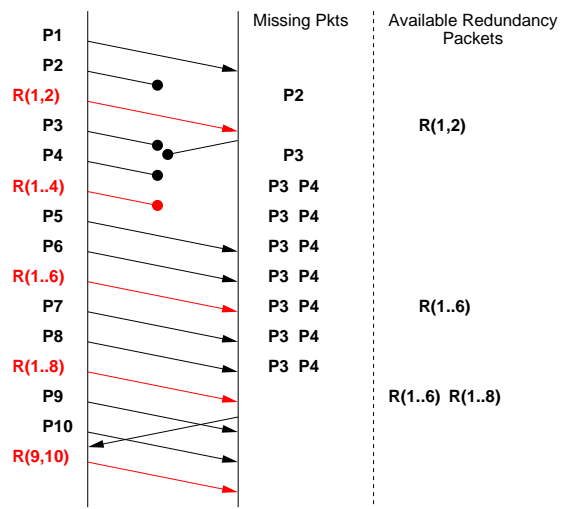}
\caption{A simple data exchange with Tetrys ($k=2$) \cite{Tetrys}}
\label{fig:tetrys_example}
\end{center}
\end{figure}

Figure \ref{fig:tetrys_example} shows a simple Tetrys data exchange with $k=2$ which implies that a repair packet is sent for every two sent source packets (equivalent to a redundancy ratio of 33.3\%). The packet $P_2$ is lost during the data exchange. However, the reception of repair packet $R_{(1,2)}$ allows the reconstruction of $P_2$. When the acknowledgment event occurs, Tetrys receiver sends a Tetrys acknowledgment packet that acknowledges packets $P_1$ and $P_2$. However, if this acknowledgment packet is lost, this loss does not interrupt the transmission; the sender simply continues to compute the repair packets from $P_1$. Later, the lost packets $P_3$, $P_4$ are reconstructed thanks to $R_{(1..6)}$ and $R_{(1..8)}$. It must be noted that the reception of packet $R_{(1..6)}$ does not allow the recovery of the first lost packet observed (packet $P_3$) since last packet recovery event (the reception of packet $R_{(1,2)}$). Indeed, the packet $P_4$ is still missing from the linear combination in packet $R_{(1..6)}$. The reception of a second acknowledgment packet allows the sender to remove the acknowledged source packets and build the repair packets from $P_9$. The reader is referred to \cite{Tetrys} for further details. 

This example deserves two important remarks. First, all lost packets (the first lost packet since the last packet recovery event as well as the last lost packet observed) are recovered altogether. Indeed, the Tetrys receiver has to wait until the number of repair packets is equal to the number of lost packets. Second, a higher redundancy ratio for the Tetrys sender leads to less delay recovery time for lost packets since the inter-arrival time between two consecutive repair packets is shortened.

\section{Redundancy adaptation algorithm for real-time video transmission}
\label{sec:adaptive_tetrys}

This section first introduces our previous work which investigated the model on packet recovery delay. Then, we present a redundancy adaptation algorithm for real-time video transmission which adapts to network dynamics based on insights from previous work.

\subsection{Previous work}
In \cite{Tetrys}, Tournoux et al. proposed a heuristic model $\theta(t)_{(d,p,b,T,R)}$ (see the notations in Table \ref{tab:notation}) for multimedia applications that requires an arrival of a certain amount of packets within a tolerable delay constraint $D_{max}$. This model gives the cumulative distribution function of lost packets recovery delay. The model assumes a Constant Bit Rate (CBR) with the same packet size that produces a data packet every $T$ seconds based on a network state (e.g., a delay $d$, a packet loss rate $p$ and a burstiness of losses $b$). The authors found that $\theta(t)_{(d,p,b,T,R)}$ fits well to the Weibull distribution which is defined by the scale $\lambda$ and the shape $\kappa$ parameters as follows:
\begin{equation}
P[X<x]=1-e^{-(x/\lambda)^\kappa}
\label{eq:weibull}
\end{equation}
\noindent where $\lambda({\Delta}_{R})$ is inversely proportional to  ${\Delta}_{R}$ and is expressed as $\lambda({\Delta}_{R})=\frac{a_{\lambda}}{{{\Delta}_{R}}^{b_{\lambda}}}$. While $\kappa({\Delta}_{R})$ evolves linearly as a function of ${\Delta}_{R}$ and is expressed as $\kappa({\Delta}_{R})=a_{\kappa}*{\Delta}_{R}+b_{\kappa}$. The coefficients $a_i$, $b_i$ ($i \in \{\lambda,\kappa$\}) are related to the loss pattern ($p$ and $b$) and $n$. However, this heuristic model has some drawbacks. First, it requires an accurate channel estimation which is not an obvious task. Furthermore, this model does not adapt well to network changes where both the loss rate, the burstiness of losses and propagation delay vary over time. However, this model does give us some insight designing a redundancy adaptation algorithm presented in section \ref{sec:algorithm}

\begin{table}[htb!]
\begin{center}
\begin{small}
\begin{tabular}{|l|m{0.8\columnwidth}|}
\hline
$k$ & The number of sent source packets between two consecutive repair packets \tabularnewline
\hline
$n$ & The total number of source packets plus a repair packet $n=k+1$ \tabularnewline
\hline
$R$ & Redundancy ratio $R=\frac{1}{n}$ \tabularnewline
\hline
$p$ & Packet loss rate \tabularnewline
\hline
$b$ & Average length of consecutive lost packets (mean burst size) \tabularnewline
\hline
${\Delta}_R$ & The difference between redundancy ratio and packet loss rate ${\Delta}_R=R-p=\frac{1}{n}-p$ \tabularnewline
\hline
$d$ & The propagation delay \tabularnewline
\hline
$D_{max}$ & The maximum tolerable delay required by the application \tabularnewline
\hline
$T$ & The mean interval time between two consecutive source packets \tabularnewline
\hline
$I$ & The mean interval time between two consecutive repair packets \tabularnewline 
\hline
$y$ &  The number of lost packets needed to be recovered in the receiver buffer\tabularnewline
\hline
$z$ &  The number of repair packets received at the receiver\tabularnewline
\hline
$Z$ & The number of additional repair packets needed to recover all losses $Z=y-z$\tabularnewline
\hline
$P_i$ & The first lost packet which has not been recovered yet since last packet recovery event \tabularnewline 
\hline
$t_i$ & The remaining time to recover the first lost packet (as well as all lost packets) before the deadline $D_{max}$ \tabularnewline 
\hline
\end{tabular} 
\end{small}
\end{center}
\caption{Notations}
\label{tab:notation}
\end{table}

\subsection{Redundancy adaptation algorithm}
\label{sec:algorithm}
\begin{figure*}[!htb]
\begin{center}
\includegraphics[scale=0.4]{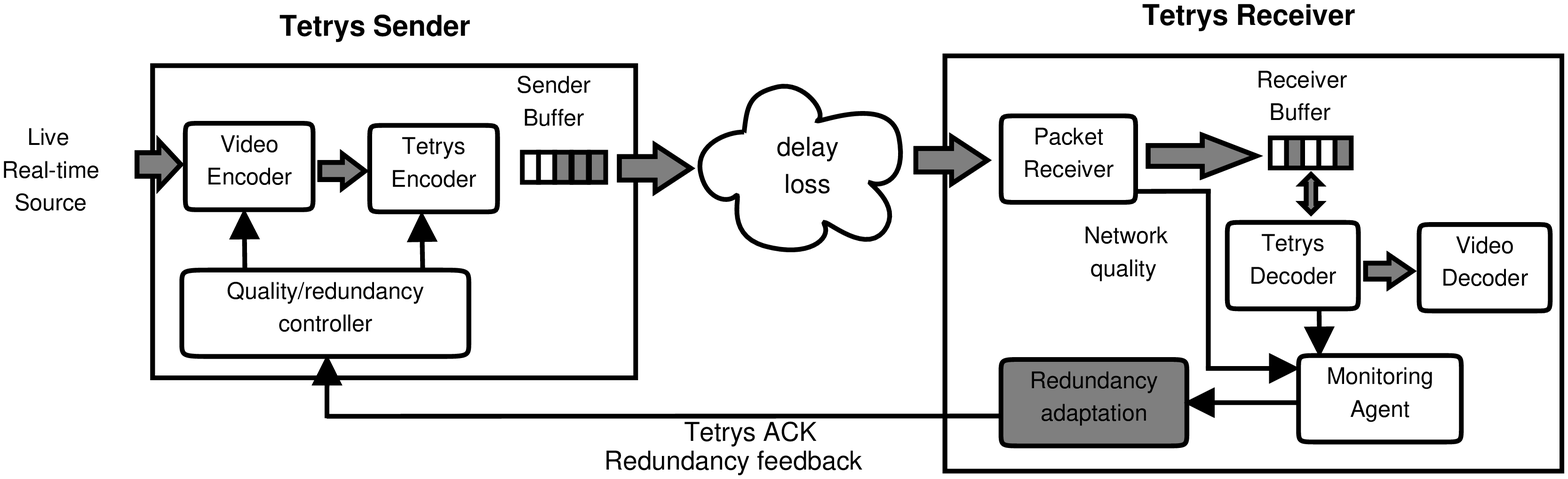}
\caption{Tetrys redundancy adaptation framework}
\label{fig:adaptation_framework}
\end{center}
\end{figure*}

The Tetrys redundancy adaptation algorithm aims to minimize the impact of packet losses in the context of real-time video transmission. Indeed, the algorithm seeks to answer the two following questions: 1) Which criteria are necessary to increase redundancy? 2) Which criteria are used to decrease redundancy?. Before answering these questions, we give an overall view of the Tetrys redundancy adaptation framework shown in Fig. \ref{fig:adaptation_framework} for real-time video transmission. The video encoder encodes the live source video based on the quality/redundancy controller. Tetrys encoder takes the encoded video and creates linear combinations for the repair packets according to the current redundancy ratio. The Tetrys receiver tries to decode the lost packets and pass the recovered lost packets to the video decoder as soon as possible. The monitoring agent observes the loss pattern and delay induced by the network. The Tetrys redundancy adaptation module gathers the information from the monitoring agent and sends increasing redundancy feedback, decreasing redundancy feedback or does nothing according to the algorithm presented below. Once the sender receives the feedback information, it changes both the redundancy ratio and video quality accordingly.

\subsubsection{Which criteria are necessary to increase redundancy?}
In section \ref{sec:tetrys_overview}, we noted that the first lost packet (as well as all lost packets) can be recovered when $Z=0$. This means that the number of received repair packets is equal to the number of lost packets. When the Tetrys receiver observes some lost packets that have not been recovered yet (or $Z>0$), it estimates the arrival time of the first lost packet $P_i$ in the absence of losses based on $T$ and the arrival time of the successfully received previous packet $P_{i-1}$. The Tetrys receiver then deduces the remaining time $t_i$ to recover packet $P_i$ as well as all lost packets before the deadline $D_{max}$ from the estimated arrival time of the packet $P_i$ and $D_{max}$. In an ideal case where there are no further losses for both data and repair packets, the Tetrys receiver needs $Z*I$ (in time) to recover all losses. The condition $Z*I < t_i$ implies that all losses can be recovered before the application constraint $D_{max}$ while $Z*I > t_i$ implies that some lost packets cannot be recovered before the application deadline. However, the algorithm actually needs $Y \geq Z$ to recover all losses, since losses may still occur up until the time when the receiver receives enough Tetrys repair packets. $Y$ depends on the loss distribution (e.g., Bernoulli or Gilbert-Elliott \cite{Frossard01}). In \cite{Tetrys}, Tournoux et al. theoretically calculate the decoding delay knowing $Z$ for the case of Bernoulli where the losses are uniformly distributed. However, this implementation is far from being trivial. Furthermore, there are no theoretical estimations of the decoding delay for other loss patterns (e.g. Gilbert-Elliott). Thus, we propose building an algorithm that increases the redundancy ratio if one of the two following conditions is not satisfied:
\begin{enumerate} 
\item $Z*I*f < t_i$
\item $P[X< t_i] \geq min_{th}$
\end{enumerate}
\noindent where $f>1$ is a coefficient that indicates the proactive level of the algorithm. The greater f means that the algorithm is more proactive to react to packet losses by adapting quickly to the redundancy before passing the application delay constraint and vice versa. The first condition implies a reactive behavior that the receiver actually observes at a given time, while the second condition indicates an estimation behavior that might occur in the future. In fact, given $t_i$, $p$ and $b$ observed at the receiver, the algorithm increases the redundancy if the probability from the Weibull function in equation (\ref{eq:weibull}) to recover the lost packets before $t_i$ is lower than a certain threshold $min_{th}$ (e.g., 0.9) which is required for the applications. When one of the two conditions is not satisfied the Tetrys receiver sends a feedback message to the Tetrys sender to require a redundancy increment.

\subsubsection{Which criteria are used to decrease redundancy?}
The Tetrys receiver sends a feedback message that requires a redundancy decrement if both of the following conditions are satisfied:
\begin{enumerate} 
\item $Z=0$
\item $P[X< D_{max}] \geq max_{th}$ 
\end{enumerate}
The first condition means that at a given time, there are no unrecovered packets. The second condition indicates that with the current redundancy ratio and the observed network state, the probability from the Weibull function of recovering packet losses before the application deadline $D_{max}$ is greater than a certain threshold $max_{th}$ (e.g., 0.99). Thus, these two conditions allow the safe reduction of redundancy for better video quality. It is clear that a given $max_{th}$ must be greater than $min_{th}$. The impact of the difference between $min_{th}$ and $max_{th}$ is studied in Section \ref{sec:impact_parameters}.

\subsection{Feedback information in Tetrys acknowledgment}
\label{sec:algo_feedback}

According to the algorithm, the Tetrys receiver sends a feedback message each time it requires a redundancy increment or decrement. These feedback messages might be lost during transmission. The loss of feedback messages that requires a redundancy decrement does not have much impact on the residual loss rate since the Tetrys sender uses a much higher redundancy ratio than the current loss rate. However, the loss of feedback messages that requires a redundancy increment has a stronger impact on the performance since the Tetrys receiver experiences the packet losses that might not be recovered before the application constraint. Furthermore, the losses may still persist or even become worse. This may lead to more lost packets that cannot be recovered before the application deadline. In a case where all increasing redundancy feedback messages are lost, the Tetrys redundancy adaptation algorithm has the same performance in terms of residual loss rate as Tetrys without the redundancy adaptation algorithm where the redundancy ratio is not changed regardless of network conditions. Thus, we propose a simple mechanism which is more robust to feedback losses. Indeed, in the event the Tetrys receiver decides to send a feedback message (redundancy increment or decrement), it sends a Tetrys acknowledgment packet in which the feedback information is included. This feedback information is also included in the periodic Tetrys acknowledgment packets afterwards until the Tetrys sender updates its redundancy ratio. Tetrys sender only updates its redundancy once when it first sees the update requirement. In this way, Tetrys does not need to handle a new packet type.

\section{Redundancy list for H.264/AVC real-time transmission}
\label{sec:correlation}
\begin{figure}[!htb]
\begin{center}
\subfigure[Encoding bit rate for different video sequences\label{fig:correlation3_a}]{\includegraphics[scale=0.33]{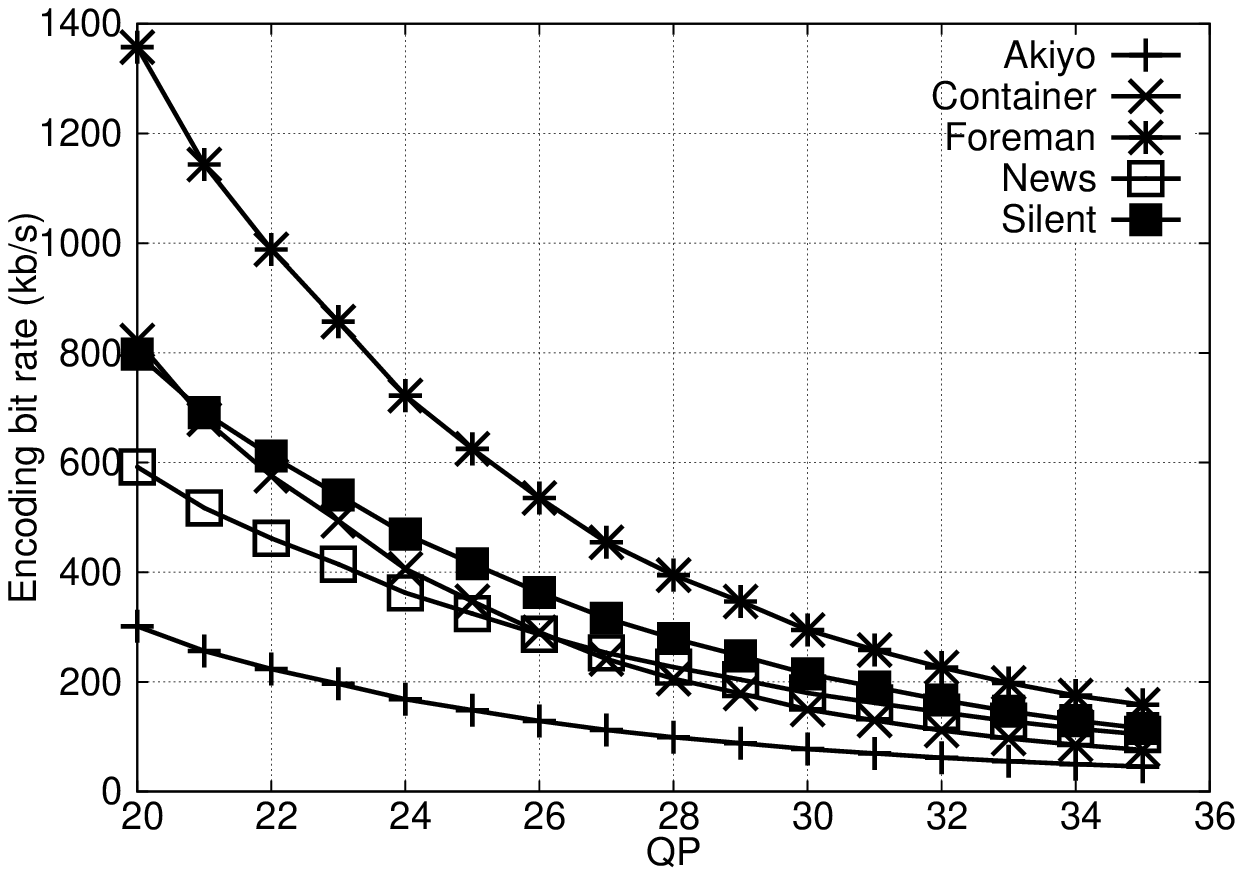}}
\subfigure[PSNR for different video sequences\label{fig:correlation3_b}]{\includegraphics[scale=0.33]{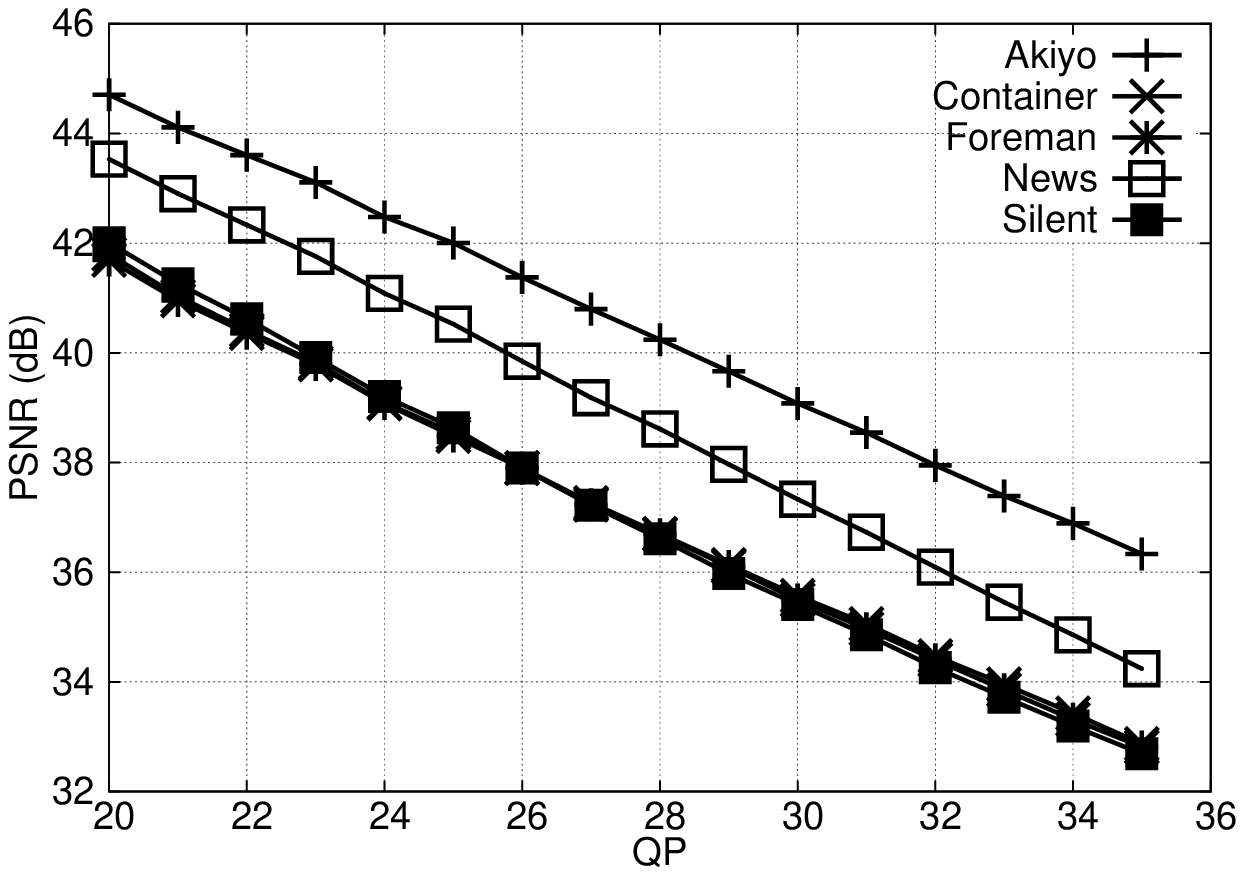}}
\subfigure[Encoding bit rate gain for 'Foreman' sequence\label{fig:correlation3_c}]{\includegraphics[scale=0.33]{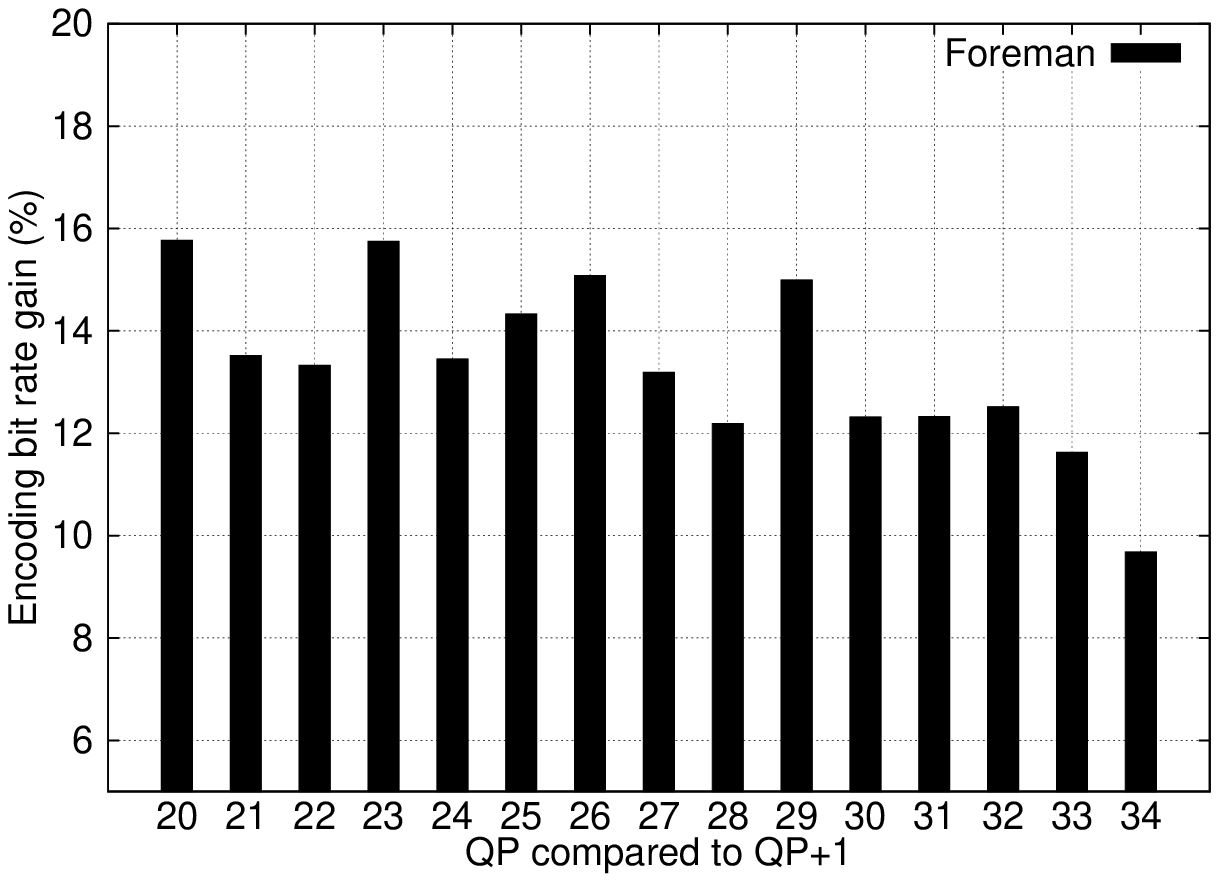}}
\subfigure[PSNR loss for 'Foreman' sequence\label{fig:correlation3_d}]{\includegraphics[scale=0.33]{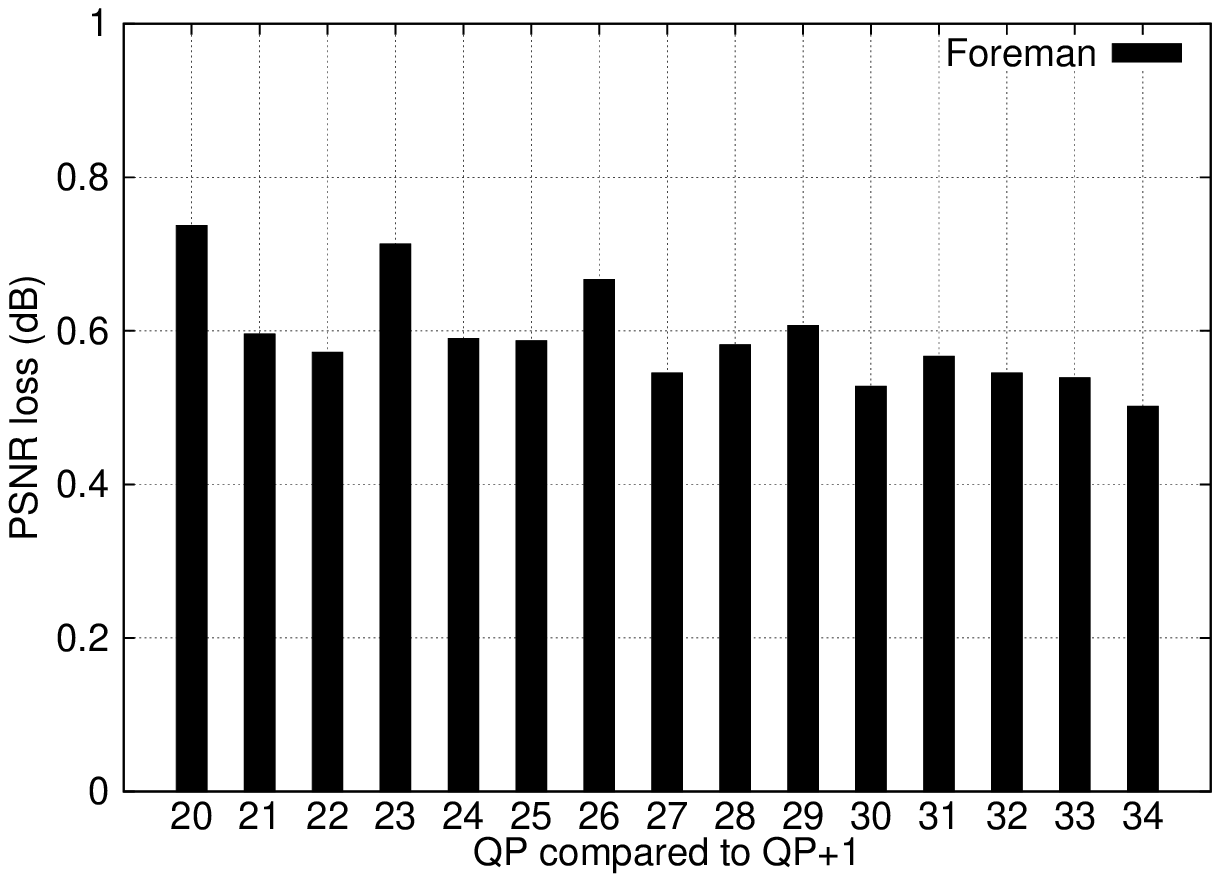}}

\caption{Different CIF video sequences encoded by x264 with Baseline profile}
\label{fig:correlation3}
\end{center}
\end{figure}

The redundancy adaptation algorithm in Section \ref{sec:algorithm} does not specify the amount of redundancy adjustment. In general, the $n$ parameter of Tetrys only takes integer values, the list of redundancy ratios is $R \in \{0.5, 0.33, 0.25, 0.2, 0.17, ...\}$ which is equivalent to the list for $n \in \{2, 3, 4, 5, 6, ...\}$. However, this general redundancy list may not fit well to video transmission where the video characteristics are taken into account. In video coding, the quantization parameter (QP) controls the trade-off between compression efficiency and image quality \cite{Richardson}. Indeed, the QP is inversely proportional to the image quality. Each time the value of QP is increased by one, the video quality is slightly degraded and vice versa. This degraded video quality comes from a lower encoding bit rate. The extra bit rate gained from lowering video quality can be used by the erasure codes to protect against packet losses. In order to estimate the variations of video bit rate due to the variations of the QP, we performed several tests with an x264 \cite{x264} encoder using CIF video format. The reference video sequences are encoded with H.264/AVC Baseline profile using $QP =QP_I = QP_P - 2 = QP_B -2$ \cite{Bergot}. Fig. \ref{fig:correlation3_a} and \ref{fig:correlation3_b} show the encoding bit rate in $kb/s$ and the PSNR in dB as a function of $QP$ for different reference video sequences. Fig. \ref{fig:correlation3_c} shows the encoding bit rate gain for the 'Foreman' sequence each time the QP is increased by one. For example, the value 20 in the $x$ axis in Fig. \ref{fig:correlation3_c} implies that when the QP is increased from 20 to 21, the bit rate gain is 15.8\% whereas the encoding bit rate decreases from $1357.2\,kb/s$ at $QP=20$ to $1143.2\,kb/s$ at $QP=21$. Similarly, Fig. \ref{fig:correlation3_d} shows the quality degradation in dB each time QP is increased by one. Similar observation on the gain in encoding bit rate and the loss in PSNR each time QP is increased by one is also obtained with the 'Akiyo', 'Container', 'News' and 'Silent' sequences. An important remark deduced from Fig. \ref{fig:correlation3} is that each time the value of QP is increased by one, the encoding bit rate gain is in the range of 10\% to 20\% while the video quality degradation is in the range of 0.5 to $1\,$dB. This percentage gain in bit rate can be used by the erasure codes to protect the video from losses. It should be noted that the impact of a slightly degraded video ranging from 0.5 to $1\,$dB is negligible to the human eye. Similar results with different video encoders (e.g., JM \cite{jm}, x264), video profiles (e.g., Baseline, High), video formats (e.g., QCIF, 4CIF, 720p), Group of Pictures (GOP) sizes and $QP$ patterns can be found in \cite{tetrys_website}. Thus, we propose a redundancy list for the case of H.264/AVC video transmission $R \in \{0.1, 0.2, 0.33, 0.5\}$ which is equivalent to the list for $n \in \{10, 5, 3, 2\}$. The chosen list of redundancy ratios ensures that the lower quality video plus redundancy used by Tetrys does not send extra bit rate compared to the normal quality video without protection. This prevents the possibility of congestion caused by the extra bit rate injected on to the networks. Let us give an example by assuming that the Tetrys sender is transmitting a video with $QP=29$ and a Tetrys redundancy ratio of 20\%. If the Tetrys sender receives increasing redundancy feedback, the Tetrys sender increases its redundancy ratio to 33.3\% while decreasing the video quality to $QP=30$. If the Tetrys sender receives decreasing redundancy feedback, the Tetrys sender reduces its redundancy ratio to 10\% while increasing the video quality to $QP=28$. In a case where the Tetrys sender receives decreasing redundancy feedback while its redundancy is 10\%, Tetrys sender maintains its redundancy ratio since 10\% is the lowest value in its redundancy list and it is necessary to protect the video from packet losses.

\section{Evaluation the algorithm parameters with CBR traffic}
\label{sec:evaluation_cbr}

\begin{figure*}[!htb]
\begin{center}
\subfigure[Coefficient $f$\label{fig:coefficient_f}]{\includegraphics[scale=0.44]{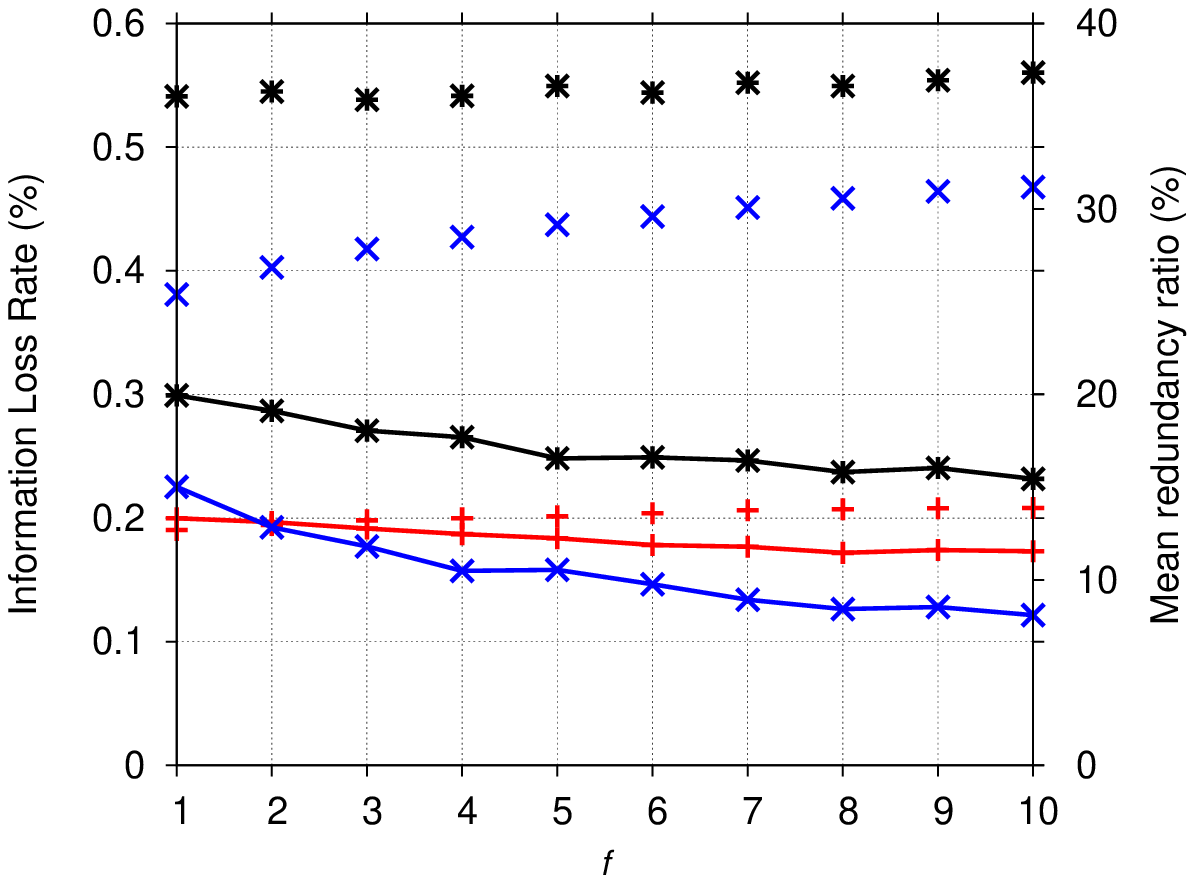}}
\subfigure[$min_{th}$\label{fig:min_threshold}]{\includegraphics[scale=0.44]{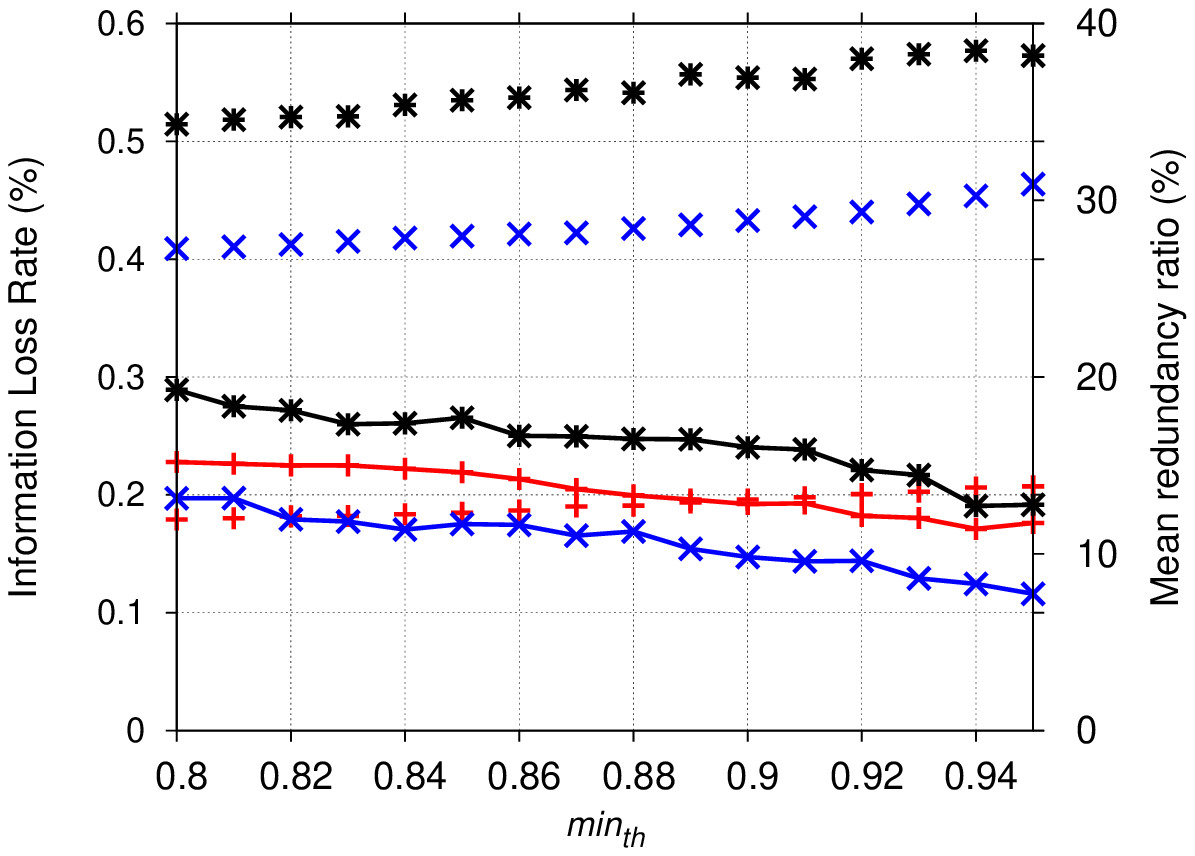}}
\subfigure[$max_{th}$\label{fig:max_threshold}]{\includegraphics[scale=0.44]{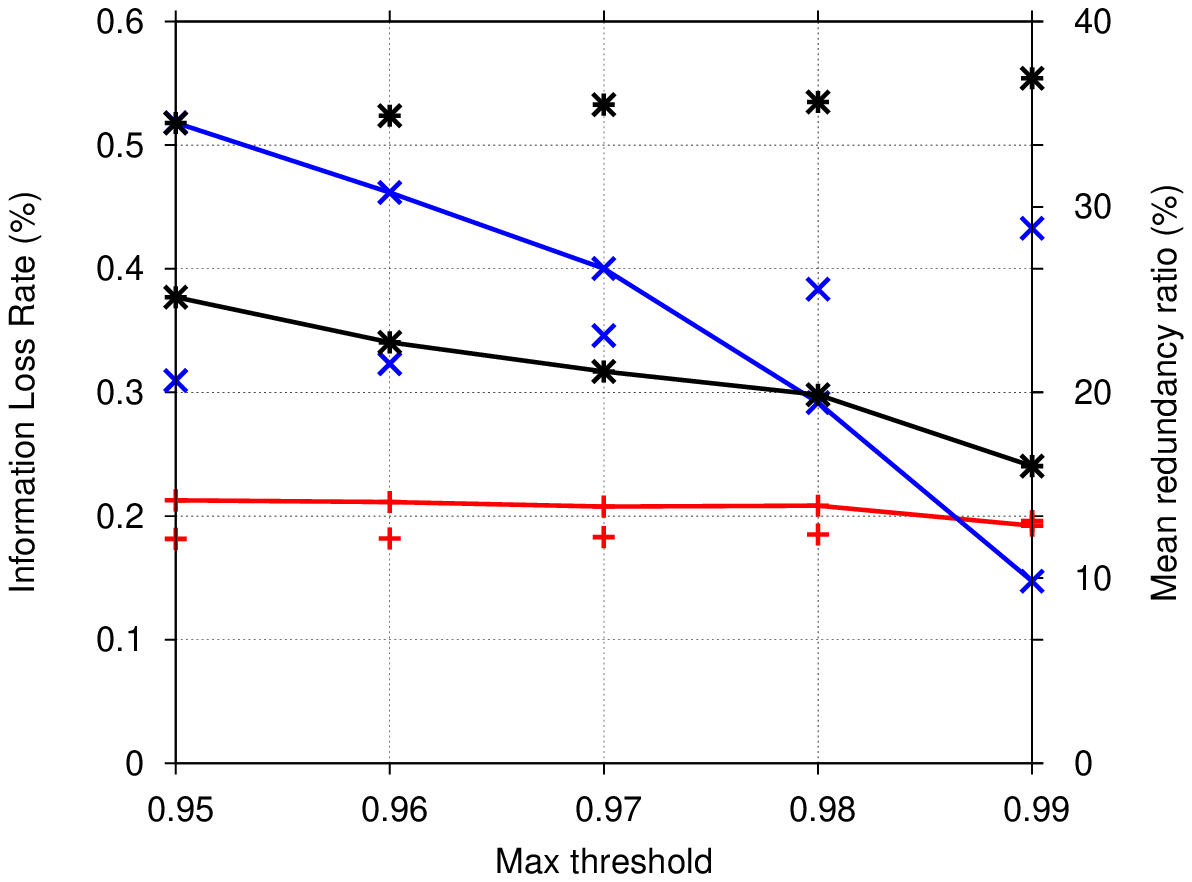}}
\caption{Impact of the algorithm parameters. The lines with dots represent the Information Loss Rate and the dots represent the average redundancy ratio. The red (+), blue (x) and black (*) colors show the PLR at 1\%, 5\% and 10\%, respectively. The channel has a mean burst size $b=3$.}
\end{center}
\end{figure*}

We evaluate the Tetrys redundancy adaptation algorithm using $ns-2$ \cite{ns}. We send a Constant Bit Rate traffic at $1900\,kb/s$ with a constant packet size of $500\, bytes$. The one-way propagation delay is set to $50\,ms$ which results in a $100\,ms$ Round Trip Time (RTT) and the one way End-to-End delay constraint $D_{max}$ is set to $150\,ms$, based on ITU-T/G.144 \cite{itu}. This constraint is recommended for highly interactive applications. The packets recovered after this deadline are considered as lost by the application. The Tetrys acknowledgment frequency is set to $10\,ms$. We evaluate the performance using an information loss rate (ILR) which indicates the residual loss rate after decoding within the application deadline at the end of each simulation. Tetrys shows best performance against uniform losses \cite{Tetrys, tetrys_multipath}, thus, we only evaluate the performance with the Gilbert-Elliott erasure channel which is specified by the average packet loss rate (PLR) and the average length of consecutive lost packets (or mean burst size) \cite{Frossard01}.  In order to provide a fair comparison, the sender sends 50000 data packets while the number of repair packets depends on the redundancy ratio used in each simulation.

\subsection{Impact of algorithm parameters}
\label{sec:impact_parameters}
We first evaluate the impact of coefficient $f$ by disabling the second condition ($P[X< t_i] \geq min_{th}$) in the increasing redundancy criteria. The $max_{th}$ is set to 0.99 in the decreasing redundancy criteria. Fig. \ref{fig:coefficient_f} shows a slight decreasing trend in ILR for different PLRs with mean burst size $b=3$ when the coefficient increases. The decrease in ILR leads to an increase in the average redundancy ratio which is shown on the second $y$ axis. The greater coefficient $f$ implies a more proactive approach against packet losses and vice versa. It is notable that the ILR of PLR = 5\% is smaller than PLR = 1\%. This can be explained by the amount of redundancy used by Tetrys in both simulations. In fact, at $f=3$, Tetrys uses on average $\approx$13\% during the simulation at PLR = 1\% while it uses on average $\approx$28\% at PLR=5\%. Then, we evaluate the impact of $min_{th}$ by disabling the first condition ($Z*I*f < t_i$) in the increasing redundancy criteria. The $max_{th}$ is still set to 0.99 in the decreasing redundancy criteria. Fig \ref{fig:min_threshold} shows that the greater value of $min_{th}$ results in a lower ILR. The remark for PLR=1\% and PLR=5\% at $b=3$ is similar to the case of coefficient $f$. Furthermore, at PLR=5\% and $b=3$, the redundancy ratio of Tetrys is greater than or equal to 20\% most of the time since the second condition in the increasing redundancy criteria is not satisfied if the redundancy ratio is 10\% compared to PLR=5\% and $b=3$. Finally, we evaluate the impact of $max_{th}$ by fixing the coefficient $f=2$ and the $min_{th}=0.9$. Fig. \ref{fig:max_threshold} shows that the algorithm suffers a higher ILR if the $max_{th}$ is low. Indeed, the lower value of $max_{th}$ implies a closer gap between $min_{th}=0.9$ and $max_{th}$ where the algorithm oscillates frequently its redundancy ratio. In fact, a redundancy ratio of 10\% is not high enough to cover a PLR of 10\%. Thus, the algorithm switches the redundancy ratio between 20\% and 33.3\% most of the time, while at PLR=5\%, the algorithm switches the redundancy ratio between 10\% and 20\%. For instance, at $max_{th}=0.95$, the redundancy ratio oscillates frequently since $max_{th}$ is close to $min_{th}=0.9$. This explains why the ILR at PLR=5\% is lower than the one at PLR=10\% when $max_{th}$ is low. Thus, we recommend using a reasonable value of $min_{th}$ that is required for applications and a high value of $max_{th}$ (> 0.98).

\subsection{Impact of losses on feedback channel}
\begin{figure}[!htb]
\begin{center}
\includegraphics[scale=0.5]{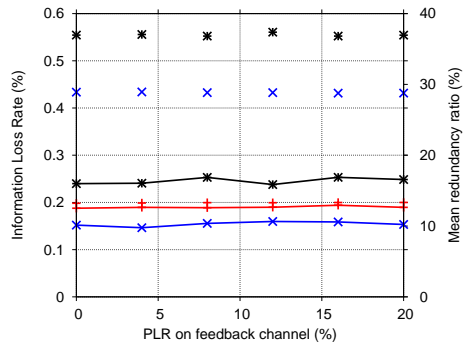}
\caption{Impact of losses on the feedback channel. The lines with dots represent the Information Loss Rate and the dots represent the average redundancy ratio. The red (+), blue (x) and black (*) colors show the PLR at 1\%, 5\% and 10\%, respectively. The forward channel has a mean burst size $b=3$. The loss rate and mean burst size on the forwarding channel is independent of the feedback channel}
\label{fig:feedback}
\end{center}
\end{figure}

To evaluate the impact of losses on the feedback channel, we conducted the same simulations as in Section \ref{sec:impact_parameters} with these settings: $f=2$, $min_{th}=0.9$ and $max_{th}=0.99$. We use a Bernoulli erasure channel for the feedback link. The loss pattern on the forwarding path is the same as previous simulations. From Fig. \ref{fig:feedback}, we see that the ILR curve is rather flat against the increasing loss rate on the feedback channel. These simulations show that the algorithm is robust to the loss rate on the feedback channel by including the feedback information in the Tetrys acknowledgment packets as presented in Section \ref{sec:algo_feedback}.

\section{Evaluation with video traffic}
\label{sec:evaluation_video}

The one-way propagation delay, the one-way E2E delay constraint and the Tetrys acknowledgment frequency are set as in Section \ref{sec:evaluation_cbr}. The 'Foreman' CIF video sequence of 300 frames is repeated 5 times to provide a video of 1500 frames at a rate of 30 frames per second. This results in 50 seconds of real-time video transmission. Thus, each 10 seconds of simulation represents a single 'Foreman' sequence. We encode the video using basic coding where there is no error resilience mechanism (e.g., Flexible Macroblock Ordering, etc.) \cite{Wenger03, KumarXMP06}. The packet size varies and depends on the encoded video. The video is encoded using the Baseline profile which is suitable for real-time video transmission. The loss concealment mechanism is frame copy. We set the coefficient $f=4$, $min_{th}=0.9$ and $max_{th}=0.99$. We evaluate the videos with three schemes: Tetrys with redundancy adaptation algorithm, Tetrys without redundancy adaptation algorithm and without Tetrys protection. The video without Tetrys protection is encoded with $QP=27$ while the video with a fixed Tetrys redundancy ratio of 10\% is encoded with $QP=28$. The QP in the video protected by Tetrys with the redundancy adaptation algorithm varies according to the redundancy ratio in such a way that the bandwidth occupation does not exceed the video without Tetrys protection. We evaluate all three scenarios. In the first scenario, the loss rate is fixed while the mean burst size varies. Both loss rate and mean burst size vary in the second scenario. Finally, both loss rate and mean burst size are fixed while the RTT varies in the third scenario.

\subsection{Evaluation with fixed loss rate and variable mean burst size}
\label{sec:eva_video_fix_plr}

\begin{figure*}[!htb]
\begin{center}
\subfigure[Tetrys with redundancy adaptation algorithm vs Tetrys without redundancy adaptation algorithm\label{fig:evaluation_fix_plr_a}]{\includegraphics[scale=0.44]{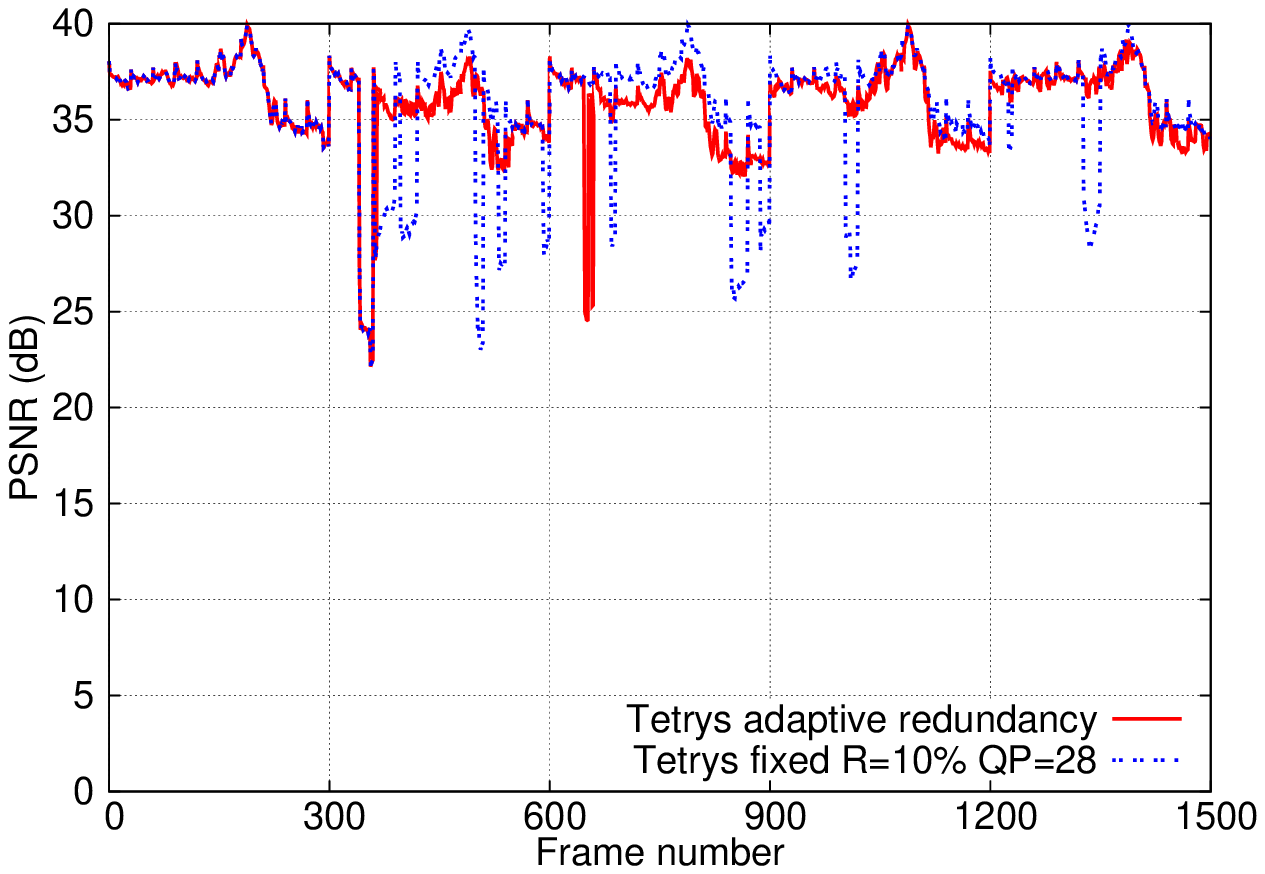}} \hspace{0.4cm}
\subfigure[Tetrys with redundancy adaptation algorithm vs without Tetrys protection\label{fig:evaluation_fix_plr_b}]{\includegraphics[scale=0.44]{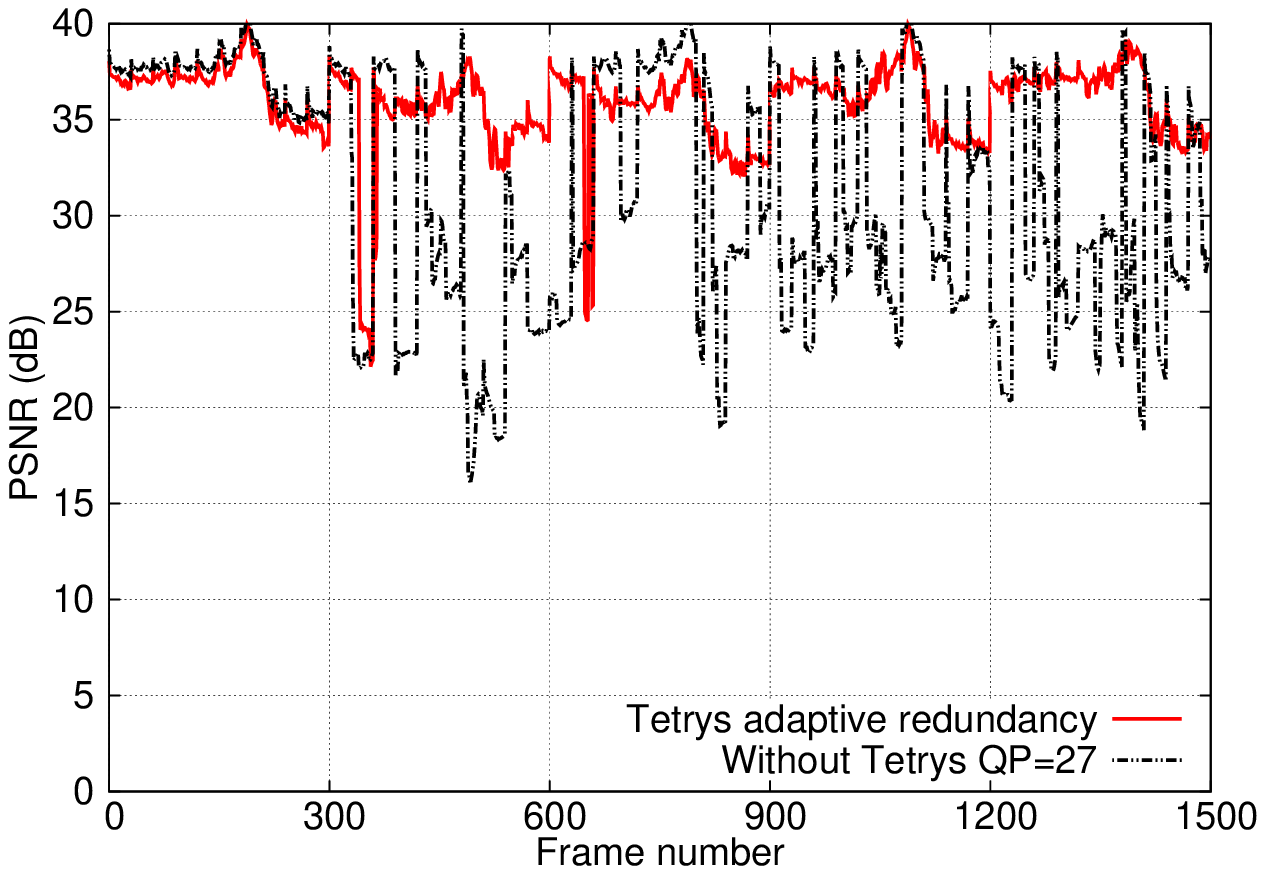}} \hspace{0.4cm}
\subfigure[Instantaneous bandwidth usage of 3 schemes \label{fig:evaluation_fix_plr_c}]{\includegraphics[scale=0.44]{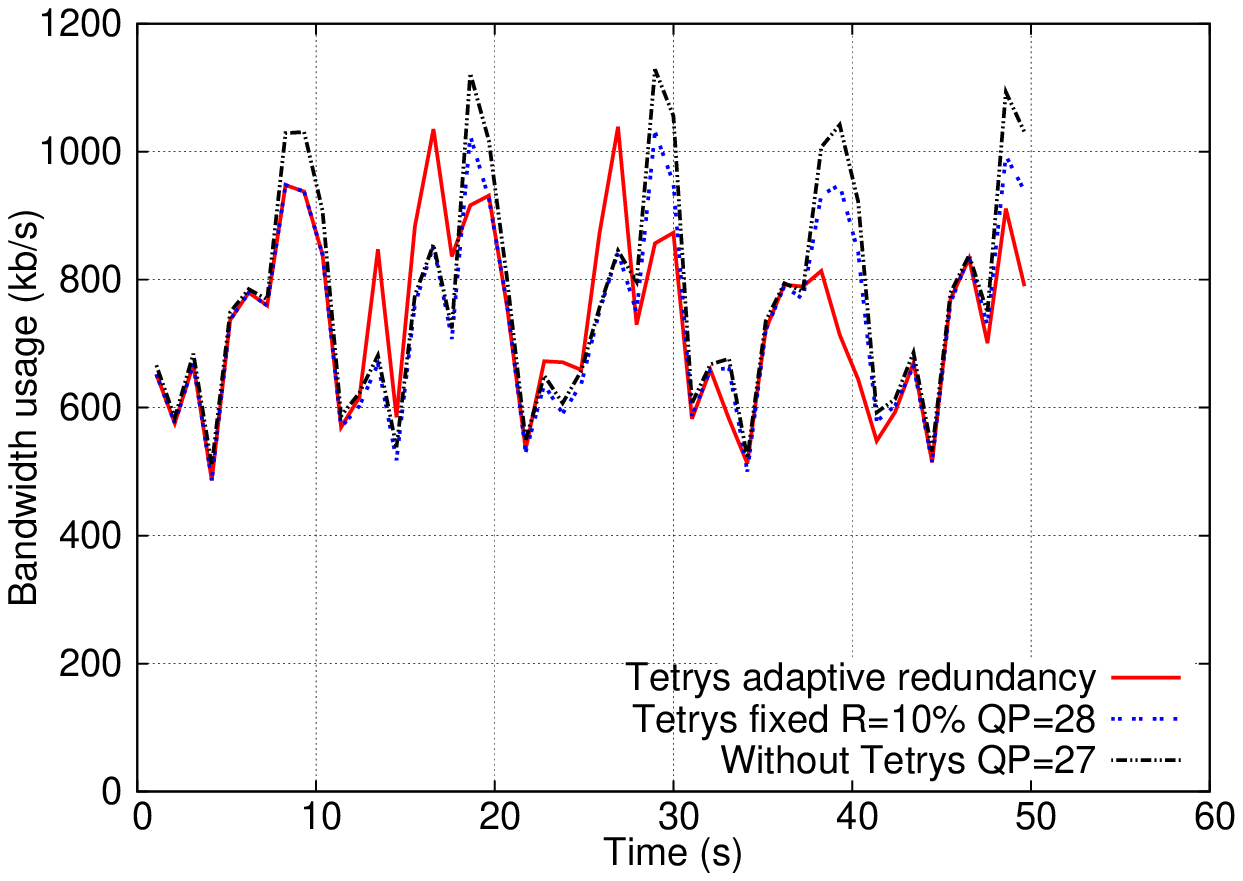}}
\caption{Comparison between 3 schemes with fixed PLR and varied $b$}
\end{center}
\end{figure*}

The loss pattern over 50 seconds of simulation is shown in Table \ref{tab:losspattern_1}. Fig. \ref{fig:evaluation_fix_plr_a} shows the results between Tetrys with adaptive redundancy and Tetrys without the redundancy adaptation algorithm. In the frame range from 0 to 300, the PSNR of Tetrys with the redundancy adaptation algorithm is the same as that of Tetrys without the redundancy adaptation algorithm since there are no losses. The Tetrys with redundancy adaptation algorithm maintains its minimum redundancy ratio of 10\%. In the frame range from 301 to 900 where the Gilbert-Elliott loss pattern with PLR=2\% and $b=2$ occurs, Tetrys without the redundancy adaptation algorithm observes a much more significant drop in quality than Tetrys with redundancy adaptation algorithm. In some frames, Tetrys with the redundancy adaptation algorithm has a slightly lower PSNR in the absence of video quality degradation. This is because Tetrys with the redundancy adaptation algorithm lowers the video quality by increasing the QP for more redundancy to adapt to network conditions. However, visually, this slightly lower quality cannot be clearly distinguished by the human eye. However, the end users suffer much stronger impact in each event where the PSNR significantly drops due to residual packet losses. In the frame range from 901 to 1500 where the Bernoulli loss pattern with PLR=2\% occurs, Tetrys without the redundancy adaptation algorithm performs well. It suffers only one quality degradation event while Tetrys with the redundancy adaptation algorithm does not experience any losses. Fig. \ref{fig:evaluation_fix_plr_b} shows the poor performance of the video without protection by Tetrys regardless of the loss pattern. Fig. \ref{fig:evaluation_fix_plr_c} shows the bandwidth usage at the outgoing interface of the sender, it can be seen that all three schemes use similar bandwidth on average. Table \ref{tab:evaluation_fix_plr} shows that Tetrys with the redundancy adaptation algorithm objectively gains on average only $0.2\,$dB compared to Tetrys without the redundancy adaptation algorithm; but subjective evaluation by watching the resulting videos \cite{tetrys_website} and Fig. \ref{fig:evaluation_fix_plr_a} shows a much better performance by Tetrys with the redundancy adaptation algorithm. Additionally, Tetrys with the redundancy adaptation algorithm and Tetrys without the redundancy adaptation algorithm both achieve the same PSNR in first 10 seconds of simulation since there are not any losses. This explains why the objective evaluation does not always adequately reflect the video quality experienced by the end users. It should be noted that the standard deviation of Tetrys with the redundancy adaptation algorithm which indicates a fluctuation in video quality is much lower than the one of Tetrys without the redundancy adaptation algorithm. Table \ref{tab:evaluation_fix_plr} also shows that the video with Tetrys redundancy adaptation algorithm uses less bandwidth on average than the video without Tetrys protection. This confirms our conservative choice of redundancy ratio list in Section \ref{sec:algorithm} where the video with the Tetrys redundancy adaptation algorithm does not use more bandwidth than the video without protection. 

\begin{table}[h!]
\caption{Loss pattern during 50s of simulation in section \ref{sec:eva_video_fix_plr}}
\begin{center}
	\begin{tabular}{|l|c|c|}
\hline
	Time (s) & Loss pattern & Frame range  \\ \hline
	0 - 10 & no losses & 0 - 300 \\ \hline
	10 - 30 & Gilbert-Elliott PLR=2\%, $b=2$ & 301 - 900 \\ \hline
	30 - 50 & Bernouilli PLR=2\% & 901 - 1500 \\ \hline
\end{tabular}
\end{center}
\label{tab:losspattern_1}
\end{table}

\begin{table}[h!]
\caption{Mean and standard deviation of psnr and bandwidth usage with different schemes in section \ref{sec:eva_video_fix_plr}}
\begin{center}
	\begin{tabular}{|l|c|c|}
\hline
	 & PSNR (dB) & BW usage (kb/s) \\ \hline
	Tetrys with adaptive redundancy& 35.9 $\pm$ 2.3 & 737.8 $\pm$ 140.3 \\ \hline
	Tetrys without adaptive redundancy & 35.7 $\pm$ 3.3 & 740.3 $\pm$ 148.7 \\ \hline
	Without Tetrys & 31.1 $\pm$ 6.4 & 774.1 $\pm$ 174.8\\ \hline
\end{tabular}
\end{center}
\label{tab:evaluation_fix_plr}
\end{table}

\subsection{Evaluation with both variable loss rate and mean burst size}
\label{sec:eva_video_varied_plr}

\begin{figure*}[!htb]
\begin{center}
\subfigure[Tetrys with redundancy adaptation algorithm vs Tetrys without redundancy adaptation algorithm\label{fig:evaluation_vary_plr_bs_a}]{\includegraphics[scale=0.44]{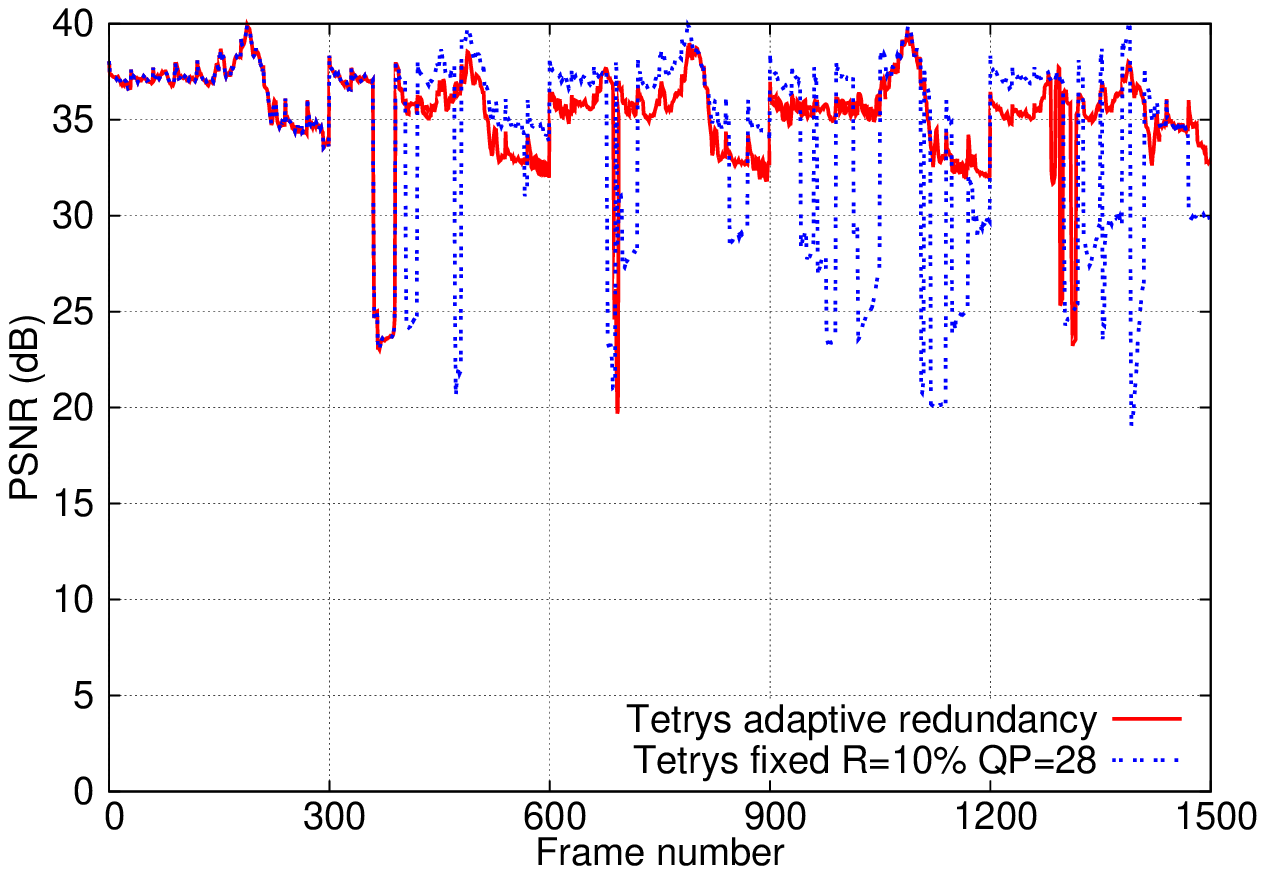}} \hspace{0.4cm}
\subfigure[Tetrys with redundancy adaptation algorithm vs without Tetrys protection\label{fig:evaluation_vary_plr_bs_b}]{\includegraphics[scale=0.44]{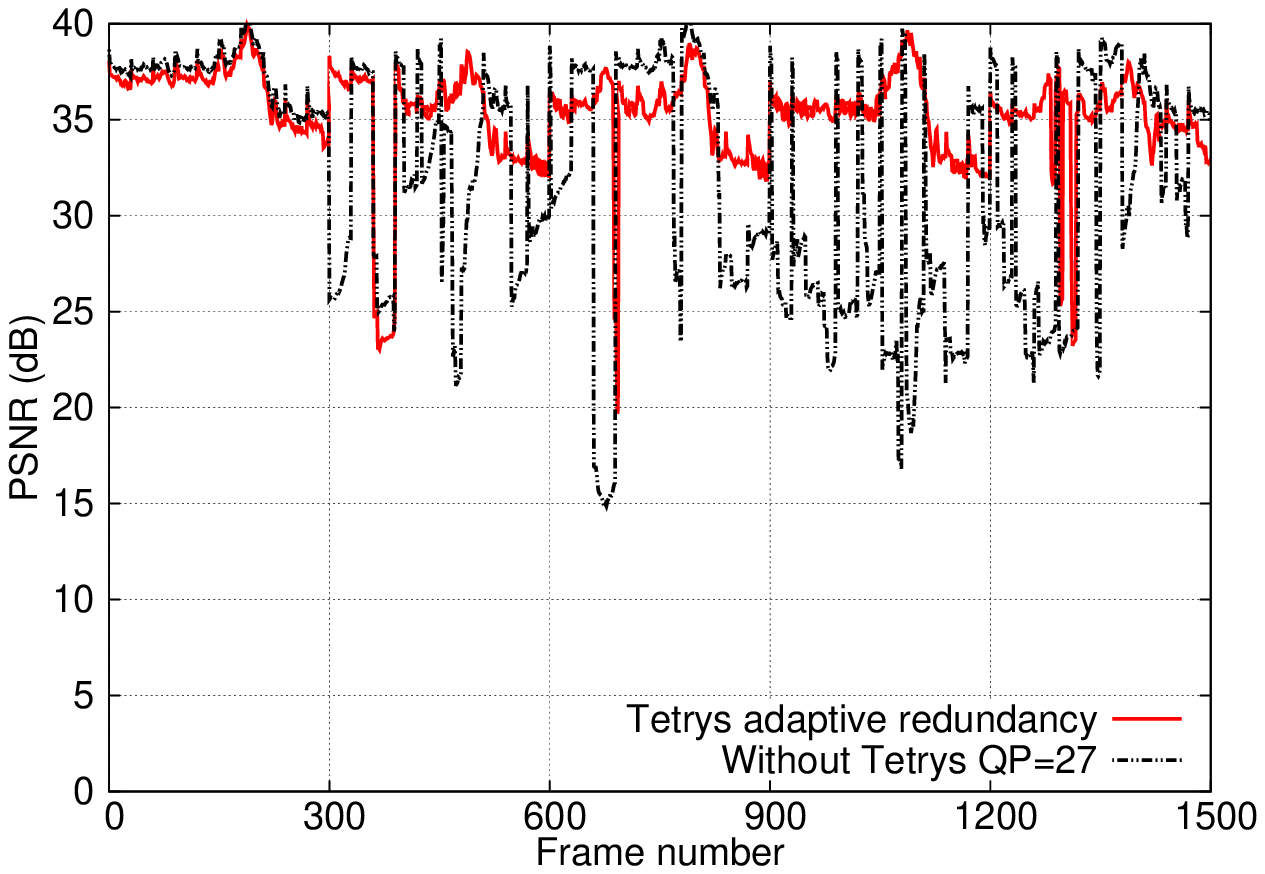}} \hspace{0.4cm}
\subfigure[Instantaneous bandwidth usage of 3 schemes\label{fig:evaluation_vary_plr_bs_c}]{\includegraphics[scale=0.44]{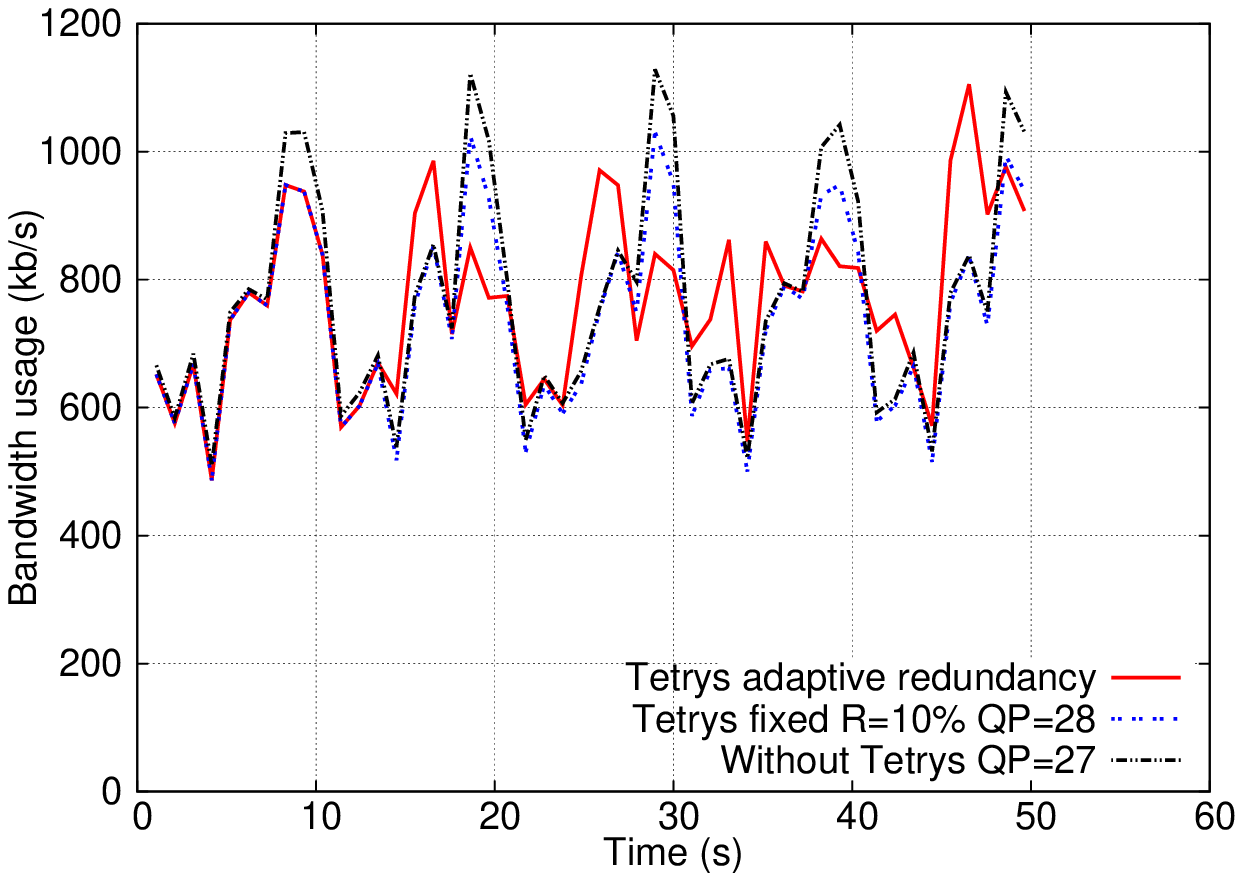}}
\caption{Comparison between 3 schemes with varied both PLR and $b$}
\end{center}
\end{figure*}

The loss pattern in this simulation is shown in Table \ref{tab:losspattern_2}. Fig. \ref{fig:evaluation_vary_plr_bs_a} shows that Tetrys without redundancy adaptation algorithm suffers from variations of both PLR and mean burst size. In fact, when the PLR is increased from 2\% to 3\% from frame 901, Tetrys without the redundancy adaptation algorithm experiences more residual losses than previous frames which leads to more video quality degradation events. Fig. \ref{fig:evaluation_vary_plr_bs_b} confirms that video without protection from erasure codes suffers poor performance from both PLR and mean burst size. The instantaneous bandwidth usage of Tetrys with the redundancy adaptation algorithm in Figure \ref{fig:evaluation_vary_plr_bs_c} is slightly different from Fig. \ref{fig:evaluation_fix_plr_c} since its uses both different redundancy ratio and video quality to adapt to the network state. Table \ref{tab:evaluation_vary_plr_bs} shows that Tetrys with the redundancy adaptation algorithm objectively gains on average $1.2\,$dB compared to Tetrys without the redundancy adaptation algorithm. Furthermore, from the subjective evaluation perspective, Tetrys with the redundancy adaptation algorithm gives a much better performance \cite{tetrys_website}. In this simulation, the video with the Tetrys redundancy adaptation algorithm uses the same average bandwidth as the video without Tetrys protection.

\begin{table}[h!]
\caption{Loss pattern during 50s of simulation in section \ref{sec:eva_video_varied_plr}}
\begin{center}
	\begin{tabular}{|l|c|c|}
\hline
	Time (s) & Loss pattern & Frame range  \\ \hline
	0 - 10 & no losses & 0 - 300 \\ \hline
	10 - 20 & Gilbert-Elliott PLR=2\%, $b=2$ & 301 - 600 \\ \hline
	20 - 30 & Gilbert-Elliott PLR=2\%, $b=3$ & 601 - 900 \\ \hline
	30 - 40 & Gilbert-Elliott PLR=3\%, $b=2$ & 901 - 1200 \\ \hline
	40 - 50 & Gilbert-Elliott PLR=3\%, $b=3$ & 1201 - 1500 \\ \hline
\end{tabular}
\end{center}
\label{tab:losspattern_2}
\end{table}

\begin{table}[h!]
\caption{Mean and standard deviation of psnr and bandwidth usage with different schemes in section \ref{sec:eva_video_varied_plr}}
\begin{center}
	\begin{tabular}{|l|c|c|}
\hline
	 & PSNR (dB) & BW usage (kb/s) \\ \hline
	Tetrys with adaptive redundancy & 35.3 $\pm$ 2.6 & 773.8.3 $\pm$ 138.1 \\ \hline
	Tetrys without adaptive redundancy & 34.1 $\pm$ 5.0 & 740.3 $\pm$ 148.7 \\ \hline
	Without Tetrys & 31.9 $\pm$ 6.2 & 774.1 $\pm$ 174.8\\ \hline
\end{tabular}
\end{center}
\label{tab:evaluation_vary_plr_bs}
\end{table}

\subsection{Evaluation with varied RTT}
\begin{figure}[htb]
    \centering
        \includegraphics[scale=0.5]{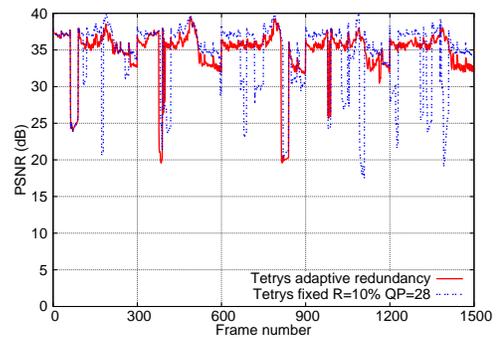}
 \caption{Tetrys with redundancy adaptation algorithm vs Tetrys without redundancy adaptation algorithm with varied RTT}
\label{fig:eva_video_vary_rtt}
\end{figure}

The loss pattern is fixed with PLR=2\% and $b=2$ during the simulation. The one-way propagation delay is set to $50\,ms$ at the beginning of the simulation and increases to $70\,ms$ after 20 seconds. Fig. \ref{fig:eva_video_vary_rtt} shows that both Tetrys with and without redundancy adaptation algorithm performs well at the one-way delay of $50\,ms$. However, when the delay is increased to $70\,ms$ where the remaining time to recover the lost packets is shortened, Tetrys without the redundancy adaptation algorithm observes a greater drop in quality. However, the performance of Tetrys with the redundancy adaptation algorithm remains constant since the algorithm takes into account this change from the signal $t_i$ and reacts accordingly.

\section{Comparison with fec adaptation scheme}
\label{sec:comparison_fec}

While Tetrys adapts to network dynamics by changing only one parameter, the redundancy ratio, redundancy adaptation with FEC is more complicated. First, FEC(k,n) which indicates $k$ source packets and $n-k$ repair packets requires changing both the group size $n$ and the redundancy ratio $(n-k)/n$. It is not evident to provide the largest group size possible before transmitting the data since FEC is more robust to burstiness losses at a larger group size. However, large group size may lead to inefficiency since FEC repair packets may arrive after the application delay constraint due to its group size or a longer delay caused by the network. Second, the criteria for adapting the FEC redundancy ratio and group size are not obvious. Tetrys has the signals from the first lost packet which has not been recovered yet and the probability of recovering the losses before the delay constraint. On the contrary, FEC must wait for the arrival of the last packet in a FEC group if it is unable to recover the lost packets with the current received packets.

In order to provide some insights into how Tetrys with the redundancy adaptation algorithm performs compared to FEC, we propose a simple redundancy algorithm for FEC with the assumption that the best FEC group size $n$ is known. The redundancy ratio list is the same as with Tetrys ([10, 20, 33.3, 50]\%). The algorithm decides to increase the redundancy if its current redundancy ratio is less than the observed loss rate plus a threshold $min_{FEC}$ mathematically presented by $R_{FEC} < p + min_{FEC}$. Similarly, the algorithm decreases the redundancy if $R_{FEC} > p + max_{FEC}$. In this case, $min_{FEC}$ must be lower than $max_{FEC}$.

\begin{figure}[!htb]
\begin{center}
\subfigure[Tetrys with adaptive redundancy algorithm vs FEC with adaptive redundancy algorithm\label{fig:evaluation_fec_a}]{\includegraphics[scale=0.5]{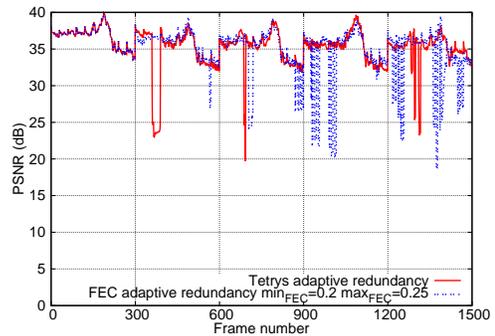}}
\subfigure[Tetrys with adaptive redundancy algorithm vs FEC with a fixed redundancy ratio of 33.3\%\label{fig:evaluation_fec_b}]{\includegraphics[scale=0.5]{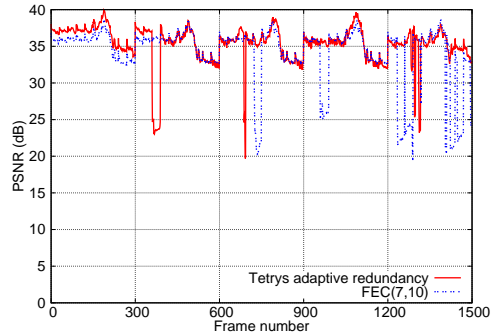}}
\caption{Tetrys vs FEC}
\end{center}
\end{figure}

We conducted several simulations to determine the largest FEC group size (i.e., the best FEC group size) that would not be inefficient. By varying the FEC group size in each simulation, we found that the largest FEC group size is 10 packets. Thus, we set $n$ at or close to 10 for the simulation and let $k$ vary according to the redundancy ratio. For instance, if the redundancy ratios are 10\% and 33.3\%, we use FEC(9,10) and FEC(6,9), respectively. It can be noted that the FEC group size can be larger with higher quality or video resolution (e.g., 4CIF or 720p) where there are more packets per image encoded than the CIF 'Foreman' video. We used the loss pattern as in Table \ref{tab:losspattern_2}. We varied $min_{FEC}$ from 0.06 to 0.2 with a step size of 0.02 and $max_{FEC}$ from 0.1 to 0.3 with a step size of 0.05 while satisfying the constraint $min_{FEC}<max_{FEC}$. We chose the combination where $max_{FEC}=0.25$ and $min_{FEC}=0.2$ that provides the best performance. The performance evaluation is based on the number of decoded frames which have a PSNR greater than $30\,$dB. Fig. \ref{fig:evaluation_fec_a} shows that at PLR=2\% with both $b=2$ and $b=3$ where the video frame ranges between 301 and 900, FEC with the adaptive redundancy achieves similar performance to Tetrys with the adaptive redundancy. However, when the PLR is increased to 3\% from frame 901, we see that FEC suffers higher video quality degradation than Tetrys. Furthermore, from frame 1201 where the mean burst size is equal to 3, FEC suffers severe quality degradation due to residual losses. Since the FEC group size is small, FEC exhibits more problems at higher burst sizes. From the simulation, FEC with the adaptive redundancy uses an average redundancy ratio of 26.2\%. To compare with traditional FEC, we also conduct a simulation with FEC(7,10) without adaptive redundancy which resulted in a redundancy ratio of 33.3\%. Even though the redundancy ratio of 33.3\% is favorable for FEC, Fig. \ref{fig:evaluation_fec_b} shows that Tetrys sill provides better performances than FEC. Table \ref{tab:evaluation_tetrys_fec} shows that Tetrys achieves a better PSNR and uses less bandwidth than FEC with and without adaptive redundancy.

Our objective is to show how the Tetrys redundancy adaptation algorithm performs relative compared to FEC. In this article, we do not provide a best FEC redundancy adaptation algorithm compared to our Tetrys redundancy adaptation algorithm. In \cite{Tetrys, tetrys_multipath}, with fixed redundancy ratio, we have shown that Tetrys outperforms FEC in both single path and multipath transmissions. As argued at the beginning of this Section, an adaptive redundancy scheme with FEC is more complicated and FEC does not have a strong enough signal to react to network dynamics.

\begin{table}[h!]
\caption{Mean and standard deviation of psnr and bandwidth usage with different schemes in section \ref{sec:comparison_fec}}
\begin{center}
	\begin{tabular}{|l|c|c|}
\hline
	 & PSNR (dB) & BW usage (kb/s) \\ \hline
	Tetrys with adaptive redundancy & 35.3 $\pm$ 2.6 & 773.8 $\pm$ 138.1 \\ \hline
	FEC with adaptive redundancy & 35.0 $\pm$ 3.5 & 889.3 $\pm$ 155.1 \\ \hline
	FEC without adaptive redundancy & 34.1 $\pm$ 4.1 & 896.5 $\pm$ 136.5 \\ \hline
\end{tabular}
\end{center}
\label{tab:evaluation_tetrys_fec}
\end{table}

\section{Related work}
\label{sec:related_work}

Our approach differs from the existing work in the following aspects. First, we use an on-the-fly and systematic erasure network coding scheme that shows better performances than FEC codes in terms of packet recovery rate in both single-path and multi-path transmissions \cite{Tetrys,tetrys_multipath}. Secondly, the Tetrys redundancy adaptation algorithm focuses on real-time video transmission with a stringent delay constraint required by applications such as video conferencing while the existing proposals target the context where the receiver has a large playout buffer \cite{RED-FEC,rate-constrained}. Lastly, our algorithm does not add extra bit rate by exploiting the relationship between the redundancy ratio and the varition of the Quantization Parameter \cite{Lee}. In \cite{EMS09}, the authors propose a FEC redundancy adaptation algorithm inside the Encoded Multipath Streaming (EMS) scheme. This algorithm increases the redundancy ratio if the residual loss rate after decoding is greater than a certain threshold and vice versa. Our approach is to minimize the residual loss rate to increase the video quality experienced by end users. Furthermore, the redundancy adjustment in \cite{EMS09} is not video-aware while our algorithm adjusts the redundancy ratio based on the changes in the Quantization Parameter.

\section{Conclusions and future work}
\label{sec:conclusion}
In this paper, we introduced a redundancy adaptation algorithm based on an on-the-fly erasure network coding scheme for real-time video transmission called Tetrys. By exploiting the relationship between the changes in the Quantization Parameter, the loss or gain in encoding bit rate and the Tetrys redundancy ratio, a video with the Tetrys redundancy adaptation algorithm achieves better video quality in terms of PSNR than both the video without the Tetrys redundancy adaptation algorithm and the video without Tetrys protection. We chose the redundancy ratio list so that the video with the Tetrys redundancy adaptation does not send more bit rate than the video without projection to prevent congestion. We have shown that the Tetrys redundancy adaptation algorithm performs well with the variations of both loss pattern and delay induced by networks. Finally, we also showed that Tetrys with the redundancy adaptation algorithm outperforms FEC with and without redundancy adaptation.

\section{Acknowledgments}
This work was supported by the French ANR grant ANR-VERS-019-02 (ARSSO project). The authors would like to thank Dr. Pierre-Ugo Tournoux for useful discussions.

\bibliographystyle{IEEEtran}
\bibliography{biblio.bib}

\end{document}